\documentclass[12pt]{article}
\usepackage{graphicx}
\input epsf.tex
\def\DESepsf(#1 width #2){\epsfxsize=#2 \epsfbox{#1}}
\def\NPB{{ Nucl. Phys.} B}

\def\PLB{{ Phys. Lett.}  B}
\def\PRL{ Phys. Rev. Lett.}
\def\PRD{{ Phys. Rev.} D}

\newsavebox{\sboxpubnumber}
\newsavebox{\sboxpubdate}
\newcommand{\pubdate}[1]{\begin{lrbox}{\sboxpubdate}{#1}\end{lrbox}}
\newcommand{\pubnumber}[1]{\begin{lrbox}{\sboxpubnumber}{\begin{tabular}{l} #1 \\
				 \usebox{\sboxpubdate}
				 \end{tabular}}
                           \end{lrbox}
                           \pubblock}
\newcommand{\Title}[1]{\begin{center} {\Large #1 } \end{center}}
\newcommand{\Author}[1]{\begin{center}{ \sc #1} \end{center}}
\newcommand{\Address}[1]{\begin{center}{ \it #1} \end{center}}

\newcommand{\pubblock}{\rightline{
			\usebox{\sboxpubnumber}}}
\newenvironment{Abstract}{\begin{quotation}  }{\end{quotation}}
\newenvironment{Presented}{\begin{quotation} \begin{center}
             PRESENTED AT\end{center}\bigskip
      \begin{center}\begin{large}}{\end{large}\end{center}
      \end{quotation}}
\newcommand{\Acknowledgements}{\bigskip  \bigskip \begin{center} \begin{large}
             \bf ACKNOWLEDGEMENTS \end{large}\end{center}}

\begin{document}

\begin{titlepage}
\pubdate{\today}                    
\pubnumber{} 

\vfill
\Title{DARK MATTER 
IN SUSY MODELS}
\vfill
\Author{R. Arnowitt, B. Dutta and Y. Santoso
}
\Address{Center For Theoretical Physics, Department of Physics, \\
Texas A$\&$M University, College Station TX 77843-4242}
\vfill
\vfill
\vfill
\begin{Abstract}
Direct detection experiments for neutralino dark matter in the Milky Way 
are examined within the framework of SUGRA models with R-parity invariance and 
grand unification at the GUT scale, $M_G$. Models of this type apply to a 
large number of phenomena, and all existing bounds on the SUSY parameter 
space due to current experimental constraints are included. For models with 
universal soft breaking at $M_G$ (mSUGRA), the Higgs mass and 
$b\rightarrow s\gamma$ 
constraints imply that the gaugino mass, $m_{1/2}$, obeys $m_{1/2} >$(300-400)GeV 
putting most of the parameter space in the co-annihilation domain where 
there is a relatively narrow band in the $m_0 - m_{1/2}$ plane. For $\mu > 0$ we 
find that the neutralino -proton cross section $\stackrel{>}{\sim} 10^{-10}$ pb for $ m_{1/2} < 1$ 
TeV, making almost all of this parameter space accessible to future planned 
detectors.  For $\mu < 0$, however, there will be large regions of parameter 
space with cross sections $< 10^{-12}$ pb, and hence unaccessible 
experimentally. If, however, the muon magnetic moment anomaly is confirmed, 
then $\mu >0$ and $m_{1/2}\stackrel{<}{\sim} 800$ GeV. Models with non-universal soft breaking in 
the third generation and Higgs sector can allow for new effects arising 
from additional early universe annihilation  through the $Z$-channel pole. 
Here cross sections that will be accessible in the near future to the next
generation of detectors can arise, and can even rise to the large values 
implied by the DAMA data. Thus dark matter detectors have the possibility 
of studying the  the post-GUT physics that control the patterns of soft 
breaking. 
 \end{Abstract}\begin{Presented}
    NON-ACCELERATOR NEW PHYSICS\\
    Dubna,Russia, \\
    June 19--23, 2001
\end{Presented}
\vfill

\end{titlepage}
\def\thefootnote{\fnsymbol{footnote}}
\setcounter{footnote}{0}

\section{Introduction}The recent BOOMERanG, Maxima and DASI data has allowed a relatively precise 
determination of the mean amount of dark matter in the universe, and these 
results are consistent with other astronomical observations. Within the 
Milky Way itself, the amount of dark matter is estimated to be
\begin{equation}\rho_{DM}\stackrel{\sim}{=}(0.3 
- 0.5){\rm GeV/cm^3}\end{equation} 
Supersymmetry with R-parity invariance possesses a natural candidate for 
cold dark matter (CDM), the lightest neutralino, $\tilde\chi^0_1$, and SUGRA 
models predict a relic density consistent with the astronomical 
observations of dark matter. Several methods for detecting the Milky Way 
neutralinos exist:

(1) Annihilation of $\tilde\chi^0_1$ in the halo of the Galaxy leading to anti-proton 
or positron signals. There have been several interesting analyses of these 
possibilities \cite{kane,gondolo}, but there are still uncertainties as to astronomical 
backgrounds.

(2) Annihilation of the $ \tilde\chi^0_1$ in the center of the Sun or Earth leading to 
neutrinos and detection of the energetic $\nu_\mu$ by neutrino telescopes 
(AMANDA, Ice Cube, ANTARES). Recent analyses \cite{barger,bottino} indicate that these 
detectors can be sensitive to such signals, but for the Minimal 
Supersymmetric Standard Model (MSSM)  one requires  
$m_{\tilde\chi^0_1} > 200$ GeV (i.e. 
$m_{1/2} > 500$ GeV) and $ \tan\beta >10$, and for 
SUGRA models one is restricted to 
$\tan\beta > 35$ \cite{barger}.

(3) Direct detection by scattering of incident $ \tilde\chi^0_1$ on nuclear targets of 
terrestrial detectors. Current detectors are sensitive to such events for 
$ \tilde\chi^0_1-p$ cross sections in the range
\begin{equation}
\sigma_{\tilde\chi^0_1-p} \stackrel{>}{\sim}
 1\times 10^{-6} {\rm pb} \end{equation}
with a possible improvement by a factor of 10 - 100 in the near future. 
Future detectors (GENIUS, Cryoarray, ZEPLIN IV) may be sensitive down to 
$(10^{-9} - 10^{-10})$ pb and we will see that this would be sufficient to 
cover the parameter space of most SUGRA models.

In the following we will consider SUGRA models with R-parity invariance 
based on grand unification at the GUT scale 
$M_G\stackrel{\sim}{=} 2\times10^{16}$ GeV. In 
particular, we will consider two classes of models: Minimal supergravity 
models (mSUGRA \cite{sugra1,sugra2}) with universal soft breaking masses at $M_G$, and 
non-universal models with non universal soft breaking at $M_G$ for the Higgs 
bosons and the third generation of squarks and sleptons .  Here the gaugino 
masses ($m_{1/2}$)  and the cubic soft breaking masses ($A_0$) at $M_G$ are assumed 
universal.

SUGRA models apply to a wide range of phenomena, and data from different 
experiments interact with each other to greatly sharpen the predictions. We 
list here the important experimental constraints:

Higgs mass: $m_h > $114 GeV \cite{higgs}. The theoretical calculation of $m_h$ still has 
an an error of $\sim3$ GeV, and so we will (conservatively) interpret this bound 
to mean $m_h(\rm theory) > 111$ GeV.

$b\rightarrow s\gamma$ branching ratio. We take a $2\sigma$ range around the central 
CLEO value \cite{bsgamma}:
\begin{equation} 1.8 \times 10^{-4} \leq B(B \rightarrow X_s\gamma) \leq 4.5 \times 
10^{-4}\end{equation}

$\tilde\chi^0_1$ relic density: We assume here

\begin{equation}
0.02\leq\Omega_{\rm DM} h^2\leq0.25\end{equation}

\noindent The lower bound takes into account of the possibility that there is more 
than one species of DM. However, results are insensitive to raising it to 
0.05 or 0.10.
\begin{figure}[htb]
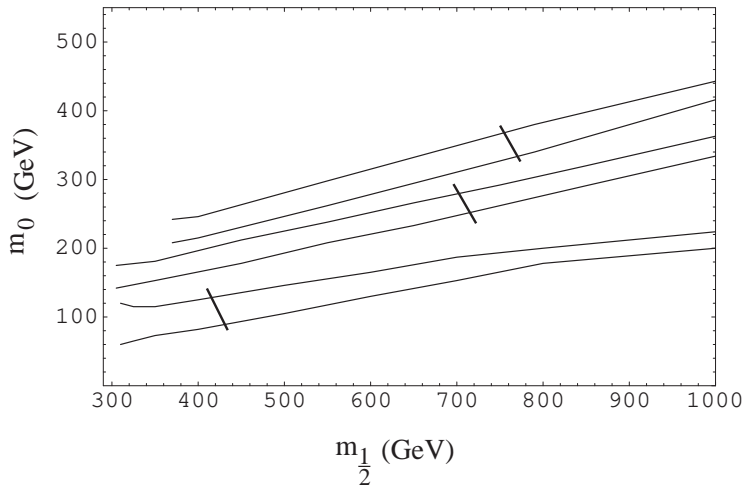

 \centerline{ \DESepsf(edmt16.epsf width 10 cm) }
\caption {\label{fig1} Corridors in the $m_0 - m_{1/2}$ plane allowed by the relic density 
constraints for (bottom to top) $\tan\beta = 10$, 30, 40, $A_0 = 0$ and $\mu > 0$. 
The lower bound on $ m_{1/2}$ is due to the $m_h$ lower bound for 
$\tan\beta$ = 10, due 
to the $b\rightarrow s \gamma$ bound for $\tan\beta =$ 40, while both these contribute 
equally for $\tan\beta = 30$. The short lines cutting the channels represent 
upper bound from the $g_\mu - 2$ experiment. [17]}
\end{figure}
\begin{figure}[htb]
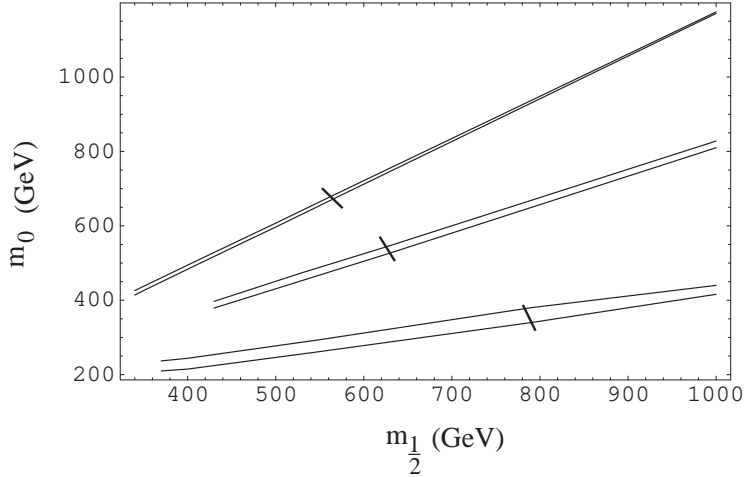

\centerline{ \DESepsf(adhs3.epsf  width 10 cm) }
\caption {\label{fig2}
 Corridors in the $m_0 - m_{1/2}$ plane allowed by the relic density 
constraint for $\tan\beta = 40,\, \mu > 0$ and (bottom to top) 
$A_0 = 0,\, -2 m_{1/2}$, 
$4m_{1/2}$. the curves terminate at the lower end due to the 
$b \rightarrow s\gamma$ 
constraint except for$ A_0 = 4 m_{1/2}$ which terminates due to the $m_h$ 
constraint. The short lines cutting the corridors represent the upper bound 
on $m_{1/2}$ due to the $g_\mu -2$ experiment. [17]}
\end{figure}

Muon $a_\mu = (g_\mu -2)/2$ anomaly. The Brookhaven E821 experiment \cite{BNL} 
reported a 2.6$\sigma$ deviation from the Standard Model value in their 
measurement of the muon magnetic moment. Recently a sign error in the 
theoretical calculation \cite{knecht,kinoshita} has reduced this to a 1.6$\sigma$ anomaly,
though recent measurements \cite{Ajinenko} used to calculate the hadronic contribution 
may have raised the deviation. Since there is a great deal of more data 
currently being analyzed (with results due this spring) that will reduce the 
errors by a factor of $\sim$2.5, we will assume here that there is a deviation 
in $a_\mu$ due to SUGRA of amount
\begin{equation}
11 \times 10^{-10}\leq a_\mu^{\rm SUGRA}\leq 75\times 10^{-10}
\end{equation}

\noindent We will, however, state our results with and without including this anomaly.

To illustrate how the different experimental constraints affect the SUSY 
parameter space, we consider the mSUGRA example:

(1) The $m_h$ and $b\rightarrow s\gamma$ constraints put a lower bound on 
$m_{1/2}$:
\begin{equation}
m_{1/2} \stackrel {>}{\sim}(300 - 400){\rm GeV}
\end{equation}
which means $m_{\tilde\chi^0_1} \stackrel {>}{\sim}(120 - 160)$ GeV (since 
$m_{\tilde\chi^0_1} \stackrel{\sim}{=} 0.4 m_{1/2}$).
(2)  Eq.(6) now  means that most of the parameter space is in the 
$\tilde\tau_1 -\tilde\chi^0_1$ co-annihilation domain in the relic density calculation. 
Then $m_0$ 
(the squark and slepton soft breaking mass) is approximately determined by 
$m_{1/2}$ as can be seen in Figs. 1 and 2.
(3) If we include the $a_{\mu}$ anomaly, since $a_{\mu}^{\rm SUGRA}$ is a decreasing 
function of $m_{1/2}$  and $m_0$ , the lower bound of Eq.(5) produces an upper 
bound on $m_{1/2}$ and the positive sign of $a_\mu$ implies that the $\mu$ parameter 
is positive. In addition one gets a lower bound on tanbeta of $\tan\beta > 5$.
\begin{figure}[htb]
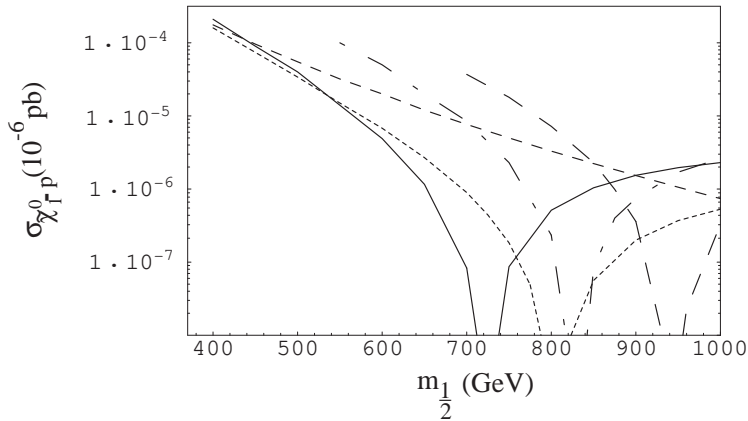

\centerline{ \DESepsf(aadcoan61020.epsf width 10 cm) }
\caption {\label{fig3} $\sigma_{\tilde{\chi}_{1}^{0}-p}$
 for mSUGRA for $\mu < 0$, $A_0 = 1500$ GeV, for $\tan\beta = 6$
(short dash), 
$\tan\beta = 8$ (dotted), $\tan\beta = 10$ (solid), $\tan\beta = 20$
(dot-dash), $\tan\beta=25$ (dashed). Note that the $\tan\beta = 6$ curve terminates
at low $m_{1/2}$ due to the Higgs mass constraint, and the other curves
terminate at low $m_{1/2}$ due to the $b \rightarrow s\gamma$ constraint [18].}
\end{figure}
\begin{figure}[htb]
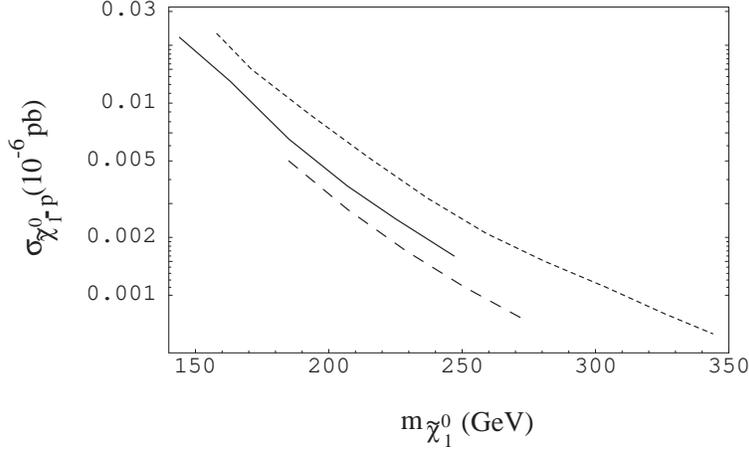

\centerline{ \DESepsf(adhs2.epsf  width 10 cm) }
\caption {\label{fig4} $\sigma_{\tilde{\chi}_1^0-p}$ as a function of the neutralino mass
$m_{\tilde{\chi}_1^0}$ for $\tan\beta = 40$, $\mu > 0$ for $A_0 = -2 m_{1/2}, 4
m_{1/2}, 0$ from bottom to top. The curves terminate 
at small $m_{\tilde{\chi}_1^0}$  due to the $b \rightarrow s\gamma$ constraint
for $A_0 = 0$ and $- 2 m_{1/2}$ and 
due to the Higgs mass bound ($m_h > 114$ GeV) for $A_0 = 4 m_{1/2}$. The curves 
terminate at large $m_{\tilde{\chi}_1^0}$ due to the lower bound on $a_{\mu}$
of Eq. (5)[17].}
\end{figure}
Thus the parameter space has begun to be strongly constrained, 
allowing for more precise predictions. In order to carry out detailed 
calculations, however, it is necessary to include a number of analyses to 
obtain accurate results. We list some of these here:

Two loop gauge and one loop Yukawa renormalization group equations (RGE) 
are used in going from $M_G$ to the electroweak weak scale $M_{\rm EW}$, and QCD RGE 
are used below $M_{\rm EW}$ for the light quark contributions.
Two loop and pole mass corrections are included in the calculation of $m_h$.
One loop corrections to $m_b$ and $m_\tau$ \cite{rattazi,carena} are included which are 
important at large $\tan\beta$.
Large $\tan\beta$  NLO SUSY corrections to $b \rightarrow s \gamma$
\cite{degrassi,carena2} are included.
In calculating the relic density, all stau-neutralino co- annihilation 
channels are included, and this calculation is done in a fashion valid for 
both small and large $\tan\beta$.

We do not include Yukawa unification or proton decay constraints, since 
these depend sensitively on post-GUT physics, about which little is known.

\section{mSUGRA MODEL}

The mSUGRA model is the simplest, and hence most predictive of the 
supergravity models in that it depends on only four new parameters and one 
sign (in addition to the usual SM parameters). We take these new 
parameters to be $m_0$ and $m_{1/2}$ (the universal soft breaking scalar and 
gaugino masses at $M_G$), $A_0$ (the universal cubic soft breaking mass at $M_G$) , 
$\tan\beta = <H_2>/<H_1>$ at the electroweak scale (where $<H_2>$ gives rise to up
 quark masses and $<H_1>$ to down quark masses) and the sign of $\mu$ (the Higgs 
mixing parameter which appears in the superpotential as $ \mu H_1 H_2$). We 
examine these parameters over the range $m_0$,$m_{1/2}\leq 1$ TeV, 
$2 < \tan\beta < 50$, $|A_0|\leq 4 m_{1/2}$. The bound on $m_{1/2}$ corresponds to the gluino mass 
bound of $m_{\tilde g} < 2.5$ GeV which is also the reach of the LHC. 

The relic density analysis involves calculating the annihilation cross 
section for neutralinos in the early universe. This characteristically 
proceeds through $Z$ and Higgs s-channel poles ($Z$, $h$, $H$, $A$ where $H$ and $A$ are 
heavy CP even and CP odd Higgs bosons) and through t-channel sfermion poles. 
However, if there is a second particle which becomes nearly degenerate 
with the neutralino, one must include it in the early universe 
annihilation processes, which then leads to the co-annihilation phenomena. 
In mSUGRA models, this accidental near degeneracy occurs naturally for the 
light stau, $\tilde\tau_1$. One can understand this semi-quantitatively by 
considering the low and intermediate $\tan\beta$ region where the RGE give for 
the right selectron, $\tilde e_R$, and the neutralino the following masses at the 
electroweak scale:
\begin{eqnarray}
 m_{\tilde e_R}^2&=& m_0^2 + 0.15 m_{1/2}^2 - sin^2\theta_W \,M_W^2 
 \cos2\beta\\
 m_{\tilde\chi^0_1}^2 &=&0.16 m_{1/2}^2                                       
\end{eqnarray}
the numerics coming from the RGE analysis. The last term in Eq. (7)
$\stackrel{\sim}{=} (40 \,{\rm GeV})^2$. Thus for $m_0 = 0$ the $\tilde e_R$ will 
become degenerate with the ${\tilde\chi^0_1}$ at 
$m_{1/2}\stackrel{\sim}{=} 400$ GeV, and co-annihilation thus begins at 
$m_{1/2}\stackrel{\sim}{=}(350 - 400 )$ 
GeV. As $m_{1/2}$ increases, $m_0$ must be raised in lock step (to keep
$m_{\tilde e_R} > 
m_{\tilde\chi^0_1}$). More precisely, it is the light stau, which is the lightest slepton 
that dominates the co-annihilation phenomena. However, one ends up with 
corridors in the $m_0 - m_{1/2}$ plane for allowed relic density with $m_0$ 
closely correlated with $m_{1/2}$ increasing as $m_{1/2}$ does, as seen in Figs. 
1 and 2. 

For dark matter detectors with heavy nuclei targets, the spin independent 
neutralino - nucleus cross section dominates, which allows one to extract 
the $\tilde\chi^0_1$ -proton cross section, $\sigma_{\tilde\chi^0_1-p}$. The basic quark diagrams for 
this scattering go through s-channel squark poles and t-channel Higgs (h, 
H) poles. The general features of  $\sigma_{\tilde\chi^0_1-p}$ that explain its properties 
are the following: 
 \begin{equation}{\rm \sigma_{\tilde\chi^0_1-p}\,\,increases  \,\,with  \,\,increasing \,\, tanbeta}      
\end{equation}\begin{equation}
{\rm \sigma_{\tilde\chi^0_1-p}\,\, decreases  \,\,with  \,\,increasing  
\,\,m_{1/2}\,\,and  \,\,increasing \,\,m_0}          \end{equation} 
Since co-annihilation generally correlates $m_0$ and $m_{1/2}$, if $m_{1/2}$ increases 
so does $m_0$ (at fixed tanbeta and $A_0$).

The smallest cross sections occur for the case $\mu < 0$. This is because a 
special cancellation can occur over a fairly wide range of $\tan\beta$ and 
$m_{1/2}$ \cite{ellis3,bdutta} driving the cross section below $10^{-13}$ pb. This is 
illustrated in Fig. 3. In these regions, there would be no hope for 
currently planned dark matter detectors to be able to detect Milky Way 
neutralinos. However, if the $a_\mu$ anomaly  is confirmed by the new  BNL 
E821 data (currently being analyzed), then $\mu < 0$ is forbidden, and the 
special cancelations do not occur for $\mu >0$. Large cross sections can then 
occur for large tanbeta. Thus is seen in Fig. 4 for $\tan\beta = 40$, with $m_h 
> 114$ GeV. If the Higgs mass bound were to rise, the lower bounds on $m_{1/2}$ 
would increase. Thus for $ m_h > 120$ GeV, one has $m_{\tilde\chi^0_1} > (200, 215, 246)$ GeV 
for $ A_0 = (-2, 0, 4 ) m_{1/2}$.

The lowest cross sections for $\mu > 0$ are expected to occur for small 
tanbeta and large $m_{1/2}$. This is seen in Fig. 5 for $\tan\beta = 10$ where one 
also sees that decreasing $A_0$ gives smaller cross sections. In general one 
finds 
\begin{equation}
  \sigma_{\tilde\chi^0_1-p}\stackrel{>}{\sim}10^{-10} \,\,{\rm pb \,\,for }\,\,
  \mu >0, \,\, m_{1/2} < 
1\,{\rm TeV} \end{equation}
Such cross sections are within the reach of future planned detectors.
\begin{figure}[htb]
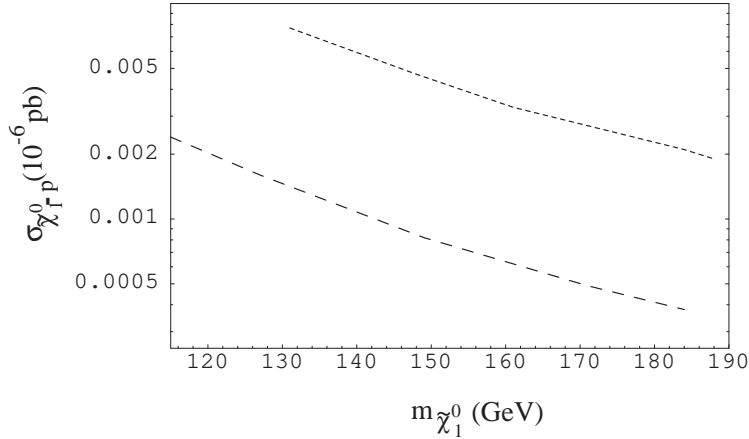

\centerline{ \DESepsf(adhs1.epsf  width 10 cm) }
\caption {\label{fig5} $\sigma_{\tilde{\chi}_1^0-p}$ as a function of $m_{\tilde{\chi}_1^0}$
for $\tan\beta = 10$, $\mu > 0$, $m_h > 114$ 
GeV for $A_0 = 0$ (upper curve), $A_0 = -4 m_{1/2}$ (lower curve). The
termination 
at low $m_{\tilde{\chi}_1^0}$ is due to the $m_h$ bound for $A_0 = 0$, and the
$b \rightarrow  s\gamma$ bound for 
$A_0 = -4 m_{1/2}$. The termination at high $m_{\tilde{\chi}_1^0}$ is due to
the lower bound on $a_{\mu}$ 
of Eq. (5)[17].}
\end{figure}
\begin{figure}[htb]
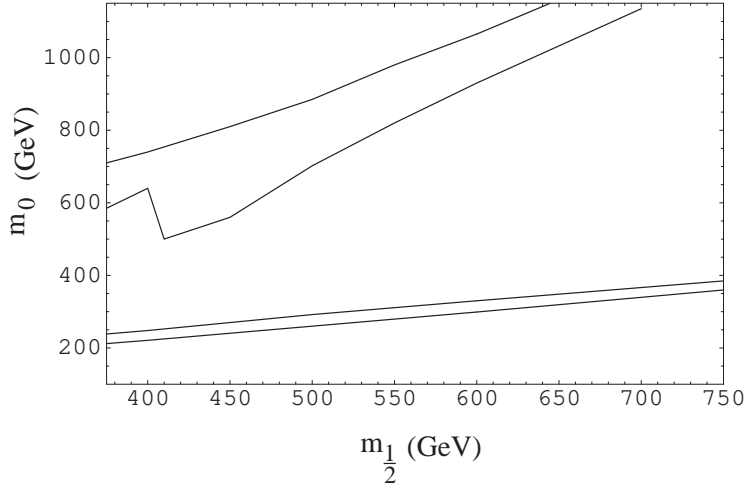

\centerline{ \DESepsf(aadcoan40newnon2.epsf width 10 cm) }
\caption {\label{fig6} Effect of a nonuniversal Higgs soft breaking mass enhancing the $Z^0$
s-channel pole contribution in the early universe annihilation, for the
case of $\delta_2 = $1, $\tan\beta = 40$, $A_0 = m_{1/2}$, $\mu > 0$. The lower band is
the usual $\tilde\tau_1$ coannihilation region. The upper band is an additional
region satisfying the relic density constraint arising from increased
annihilation via the $Z^0$ pole due to the decrease in $\mu^2$ increasing the
higgsino content of the neutralino[18].}
\end{figure}
\begin{figure}[htb]
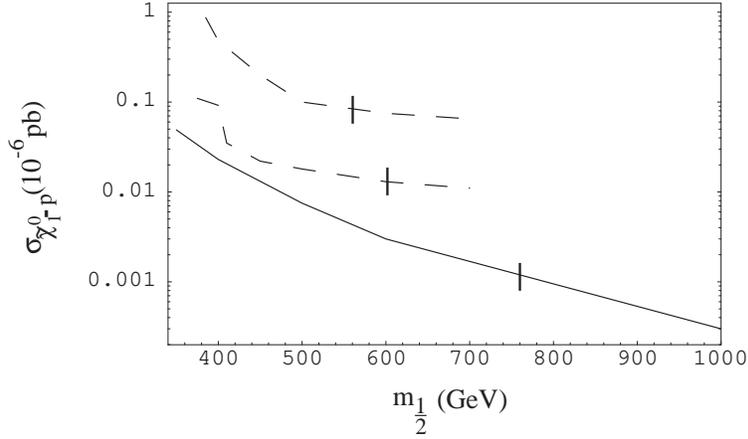

\centerline{ \DESepsf(adhs4.epsf  width 10 cm) }
\caption {\label{fig7} $\sigma_{\tilde{\chi}_1^0-p}$ as a function of $m_{1/2}$
($m_{\tilde{\chi}_1^0} \stackrel{\sim}{=} 0.4 m_{1/2}$) for $\tan\beta = 40$, 
$\mu >0$, $m_h > 114$ GeV, $A_0 = m_{1/2}$ for $\delta_2 = 1$. The lower curve
is for the $\tilde{\tau}_1-\tilde{\chi}_1^0$ co-annihilation channel, and the
dashed band is for the $Z$ s-channel 
annihilation allowed by non-universal soft breaking. The curves terminate 
at low $m_{1/2}$ due to the $b \rightarrow s\gamma$ constraint. The vertical
lines show the 
termination at high $m_{1/2}$ due to the lower bound on $a_{\mu}$ of Eq. (5)[21].}
\end{figure}

\begin{figure}[htb]
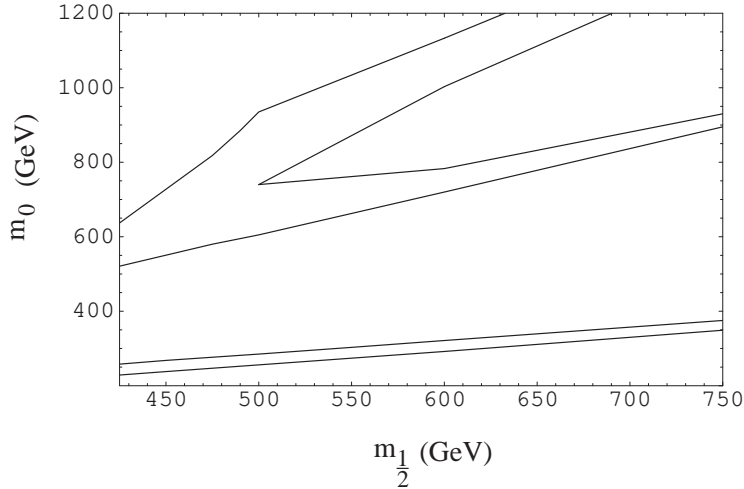

\centerline{ \DESepsf(aadcoan40newnon.epsf width 10 cm) }
\caption {\label{fig8} Allowed regions in the $m_0-m_{1/2}$ plane for the case 
$\tan\beta = 40$, $A_0
= m_{1/2}$, $\mu > 0$. The bottom curve is the mSUGRA $\tilde\tau_1$ coannihilation band of
Fig. 1 (shown for reference). The middle band is the actual $\tilde\tau_1$
coannihilation band when $\delta_{10} = -0.7$. The top band is an additional
allowed region due to the enhancement of the $Z^0$ s-channel annihilation
arising from the nonuniversality lowering the value of $\mu^2$ and hence raising the higgsino content of the neutralino. 
For $m_{1/2}\stackrel{<}{\sim}$ 500 GeV, the
two bands overlap [18].}
\end{figure}
\begin{figure}[htb]
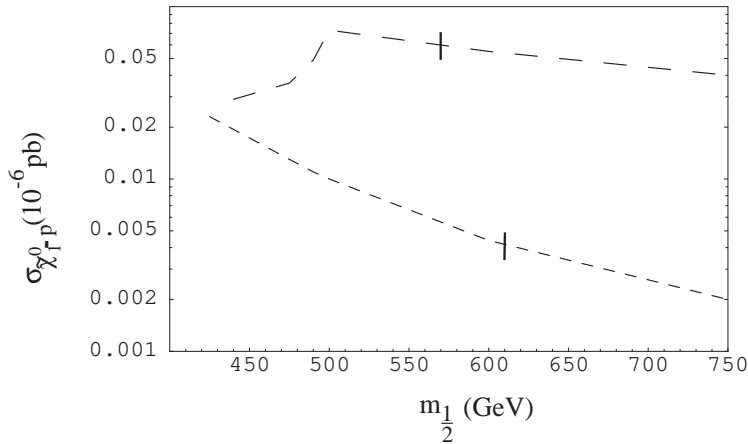

\centerline{ \DESepsf(adhs5.epsf  width 10 cm) }
\caption {\label{fig9} $\sigma_{\tilde{\chi}_1^0-p}$ as a function of $m_{1/2}$ for $\tan\beta =
40$, $\mu >0$, $A_0 = m_{1/2}$ 
and $m_h  > 114$ GeV. The lower curve is for the bottom of the
$\tilde{\tau}_1-\tilde{\chi}_1^0$ 
co-annihilation corridor, and the upper curve is for the top of the $Z$ 
channel band. The termination at low $m_{1/2}$ is due to the $b \rightarrow
s\gamma$ 
constraint, and the vertical lines are the upper bound on  $m_{1/2}$ due to the 
lower bound of $a_{\mu}$ of Eq. (5)[21].}
\end{figure}

\begin{figure}[htb]
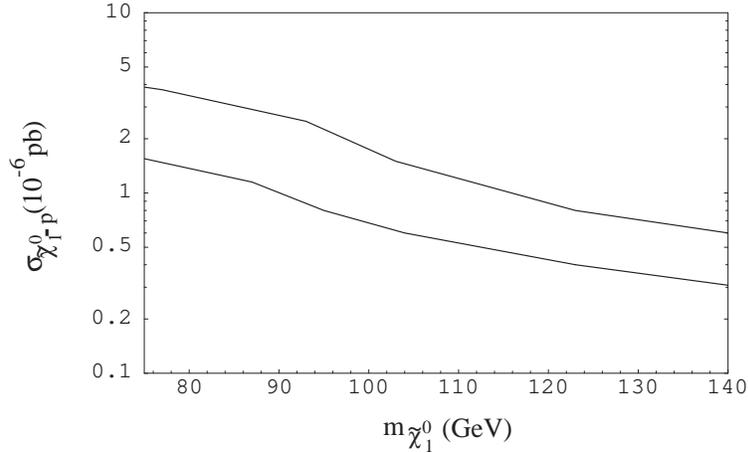

\centerline{ \DESepsf(aadtan712new.epsf width 10 cm) }
\caption {\label{fig10}Maximum value of $\sigma_{\tilde{\chi}_{1}^{0}-p}$ as a function of
$m_{\tilde{\chi}_{1}^{0}}$ for the nonuniversal
model with $\mu > 0$, $\delta_1$, $\delta_3$, $\delta_4 < 0$, $\delta_2 >0$. The lower curve
is for $\tan\beta = 7$, the upper curve is for $\tan\beta =12$.}
\end{figure}
\section{NON-UNIVERSAL MODELS}

New results can occur if we relax the universality of the squark, slepton 
and soft breaking Higgs masses at $M_G$. To maintain the flavor changing 
neutral current bounds, we do this only in the third generation and for 
the Higgs bosons. One may parameterize the soft breaking masses at $M_G$ as 
follows:
\begin{eqnarray} 
m_{H_{1}}^{\ 2}&=&m_{0}^{2}(1+\delta_{1}); 
\quad m_{H_{2}}^{\ 2}=m_{0}^{2}(1+ \delta_{2});\nonumber \\ m_{q_{L}}^{\
2}&=&m_{0}^{2}(1+\delta_{3}); \quad m_{t_{R}}^{\ 2}=m_{0}^{2}(1+\delta_{4});
\quad m_{\tau_{R}}^{\ 2}=m_{0}^{2}(1+\delta_{5});  \nonumber \\ m_{b_{R}}^{\
2}&=&m_{0}^{2}(1+\delta_{6}); \quad m_{l_{L}}^{\ 2}=m_{0}^{2}(1+\delta_{7}).
\label{eq26}
\end{eqnarray}
with $-1 \leq \delta_i \leq +1$. While the non-universal models introduce a 
number of new parameters, it is possible to understand qualitatively what 
effects they produce on dark matter detection rates, since the parameter 
$\mu^2$ governs much of the physics. Thus as $\mu^2$ decreases (increases), the 
higgsino content of the neutralino increases (decreases), and then 
$\sigma_{\tilde\chi^0_1-p}$ increases (decreases). One can further see semi-quanitatively 
the dependence of $\mu^2$ on the non-universal parameters for low and 
intermediate $\tan\beta$ where the RGE may be solved analytically \cite{nath}:
\begin{eqnarray}
\mu^2&=&{t^2\over{t^2-1}}\left[({{1-3 D_0}\over 2}+{1\over
t^2})+{{1-D_0}\over2}(\delta_3+\delta_4)\right. \nonumber \\ 
&-&\left.{{1+D_0}\over2}\delta_2+{\delta_1\over
t^2}\right]m_0^2+{\rm {universal\,parts\,+\,loop \, corrections}}. \label{eq28}
\end{eqnarray}
where $t=\tan\beta$ and $D_0 \stackrel{\sim}{=} 1 - (m_t/200\sin\beta)^2$. 
In general $D_0$ is small 
i.e. $D_0\stackrel{\sim}{=}0.25$, and one sees that the universal part of the 
$m_0^2$ 
contribution is quite small, and it does not take a great deal of 
non-universal contribution to produce additional effects.

Most interesting things happen when $\mu^2$ is decreased, since the increased 
Higgsino content of the neutralino increases the 
$\tilde\chi^0_1 - \tilde\chi^0_1- Z$ coupling, 
and this coupling opens a new annihilation channel through the $Z$-pole in 
the relic density calculations. As a simple example we consider the case 
where only the $H_2$ soft breaking mass is affected i. e. $\delta_2 = 1$ and 
all other $\delta_i = 0$. Fig. 6 shows the new allowed region in the $m_0 - 
m_{1/2}$ plane for $\tan\beta = 40$, $ A_0 = m_{1/2}$ , $\mu > 0$, and Fig. 7 shows the 
corresponding effect on the neutralino - proton cross section. On sees 
that the co-annihilation corridor is significantly raised and widened due 
to the new $Z$-channel annihilation, and the cross section is significantly 
increased. The next round of upgraded dark matter detectors should be able 
to reach parts of this parameter space if such a non-universality were to 
occur.

As a second example we consider a soft breaking pattern consistent with an 
$SU(5)$ invariant model with $\delta_{10} (= \delta_3 = \delta_4 = \delta_5) = -0.7$, 
and all other $\delta_i = 0$. Here the $\tilde\tau_R$ soft breaking mass is reduced , 
i.e $m_{\tilde \tau_R}^2 = m_0^2(1 + \delta_5) < m_0^2$. Thus the 
$\tilde\tau_1 - \tilde\chi^0_1$ 
co-annihilation occurs at a larger value of $m_0$ than in mSUGRA. In addition 
again a new $Z$-channel neutralino annihilation channel occurs since $\mu^2$ is 
reduced. The effects are shown in Figs. 8 and 9 for $\tan\beta = 40$, $A_0 = 
m_{1/2}$, $\mu > 0$. Again the cross sections are larger, and should be 
accessible to CDMS when it moves to the Soudan mine and to GENIUS.

The maximum value of $\sigma_{\tilde\chi^0_1-p}$ for fixed $\tan\beta$ and 
$A_0$ occurs when we 
chose the non-universalities to minimize $\mu^2$. This occurs  when 
$\delta_{1,3,4} < 0$ and $\delta_2 > 0$. This is shown in Fig. 10 where the 
maximum cross section is plotted for $A_0 = 0$, $\tan\beta = 12$ (upper curve), 
$\tan\beta = 7$ (lower curve). The bound that $m_h > 114$ GeV, eliminates the 
region with $m_{\tilde\chi^0_1} < 100$ GeV. However, one sees for this case that it is 
possible to have detection cross sections in the region of the DAMA data.

\section{CONCLUSIONS}

We have discussed here direct detection of Milky Way neutralinos for SUGRA 
type models with R-parity invariance and grand unification at the GUT 
scale. By combining data from a variety of sources, e.g. Higgs mass bound, 
$b\rightarrow s\gamma$ branching ratio, relic density constraints and the possible 
new muon magnetic moment anomaly  of the BNL E821 experiment, one can 
greatly sharpen predictions. 

For the mSUGRA model, the $m_h$ and $b\rightarrow s\gamma$ bounds create a lower bound on 
$m_{1/2}$ of $m_{1/2} \stackrel{>}{\sim}(300-400)$GeV (i. e. 
$m_{\tilde\chi^0_1}  \stackrel{>}{\sim} (120-140)$GeV). Thus puts the 
parameter space mostly in the $\tilde\tau_1-\tilde\chi^0_1$ co-annihilation domain, which 
strongly correlates $m_0$ with $m_{1/2}$. For $\mu > 0$ and $m_{1/2}< 1$TeV, one finds 
$\sigma_{\tilde\chi^0_1-p}\stackrel{<}{\sim} 10^{-10}$ pb which is within the upper reach of future planned 
dark matter detectors, while for $\mu < 0$ there will be large regions 
unaccessible to such detectors. If the $a_\mu$ anomaly is confirmed, then $\mu 
>0$ and $m_{1/2} < 800$ GeV.

Non-universal soft breaking models allow one to raise 
$\sigma_{\tilde\chi^0_1-p}$ by a 
factor as large as 10 - 100, which could account for the large  cross 
sections of the DAMA data. They can also open new allowed regions of the 
$m_0 - m_{1/2}$ plane from the $Z$ channel annihilation in the relic density 
calculation. The new $Z$-channel regions have larger cross sections, though 
still below the DAMA region, but they should be accessible when CDMS is in 
the SOUDAN mine and to the GENIUS-TF detector. Thus dark matter detectors 
should be able to investigate the nature of SUSY soft breaking, i.e. the 
nature of the post-GUT physics that determine the soft breaking pattern. 

\Acknowledgements This work was supported in part by National Science 
Foundation Grant PHY-0070964 and 
PHY-0101015.


\begin{thebibliography}{99}


\bibitem{kane} G.L. Kane, L.-T. Wang, J.D. Wells, hep-ph/0108138.
\bibitem{gondolo} E.A. Baltz, J.Edsjo, K. Freese,  P.Gondolo,
astro-ph/0109318.
\bibitem{barger}V.Barger, F. Halzen, D. Hooper, C. Kao, hep-ph/0105182. 
\bibitem{bottino} A. Bottino, N. Fornengo, S. Scopel, F. Donato,
hep-ph/0105233.
\bibitem{sugra1}A.H. Chamseddine, R. Arnowitt, P. Nath,
{\PRL} {\bf 49} {(1982)} {970}.
\bibitem{sugra2} R. Barbieri, S. Ferrara, C.A. Savoy,
{\PLB}{\bf 119} {(1982)} {343}; L. Hall, J. Lykken, S. Weinberg,
{\PRD}{\bf 27} {(1983)}  {(2359)}; P. Nath, R. Arnowitt, A.H. Chamseddine,
{\NPB}{\bf 227} {(1983)}   {(121)}.

\bibitem{higgs} P.~Igo-Kemenes, LEPC meeting, November 3, 2000 
(http://lephiggs.web.cern. ch/LEPHIGGS/talks/index.html). 

\bibitem{bsgamma} CLEO Collaboration, {\PRL} {\bf 87} {(2001)} {251807}.

\bibitem{BNL}H.N. Brown et.al., Muon (g-2) Collaboration, 
{\PRL} {\bf 86} {(2001)} {2227}.

\bibitem{knecht}M. Knecht, A. Nyffeler, hep-ph/0111058; 
M. Knecht, A. Nyffeler, E. De Raphael, hep-ph/0111059.

\bibitem{kinoshita}M. Hayakawa, T. Kinoshita,  hep-ph/0112102; I. Blokland, A.
Czernecki, K. Melnikov, hep-ph/0112117.
\bibitem{Ajinenko} R.R. Akhmetshin et al., hep-ex/0112031.
\bibitem{rattazi}R. Rattazi, U. Sarid, {\PRD}{\bf 53}  {(1996)} {(1553)}.
 
\bibitem{carena}M. Carena, M. Olechowski, S. Pokorski, C. Wagner
{\NPB}{\bf 426} {(1994)} {(269)}.

\bibitem{degrassi}G. Degrassi, P. Gambino, G. Giudice,
{JHEP}{\bf 0012} {(2000)} {009}.
\bibitem{carena2}M. Carena, D. Garcia, U. Nierste, C. Wagner,
{\PLB} {\bf 499} {(2001)} {141}.

\bibitem{bdutta1}R. Arnowitt, B. Dutta,   Y. Santoso,
{\PRD} {\bf 64} {(2001)} {113010}.

\bibitem{bdutta} R. Arnowitt, B. Dutta, Y. Santoso, 
{\NPB} {\bf 606}  {(2001)} {59}.

\bibitem{ellis3}J. Ellis, A. Ferstl, K. A. Olive, 
{\PLB} {\bf 481} {(2001)} {304}; {\PRD} {\bf 63} {(2001)} {065016}.

\bibitem{nath}R. Arnowitt, P. Nath, {\PRD} {\bf 56} {(1997)} {2820}.

\bibitem{bdutta2}R. Arnowitt, B. Dutta, B. Hu,  Y. Santoso,
Phys. Lett. {\bf B505},  (2001) 177.

\end{thebibliography}
\end{document}